# Ultrafast bursts of tailored spatiotemporal vortex pulses

Xin Liu[1,2,3,†], Chunhao Liang[2,3,†], Qian Cao[1,4], Yangjian Cai[2,3,*], Qiwen Zhan[1,4,5,6*]

[1] School of Optical-Electrical and Computer Engineering, University of Shanghai for Science and Technology, Shanghai 200093, China.
[2] Shandong Provincial Engineering and Technical Center of Light Manipulations and Shandong Provincial Key Laboratory of Optics and Photonic Device, School of Physics and Electronics, Shandong Normal University, Jinan 250014, China.
[3] Collaborative Innovation Center of Light Manipulations and Applications, Shandong Normal University, Jinan 250358, China.
[4] Zhangjiang Laboratory, 100 Haike Road, Shanghai 201204, China.
[5] Westlake Institute for Optoelectronics, Fuyang, Hangzhou 311421, China.
[6] International Institute for Sustainability with Knotted Chiral Meta Matter (WPI-SKCM²), Hiroshima University, Higashihiroshima, Hiroshima 739-8526, Japan.

† These authors contributed equally to this work.
* Corresponding authors: yangjiancai@sdnu.edu.cn; qwzhan@usst.edu.cn.

**Orbital angular momentums (OAMs) of light can be categorized into longitudinal OAM (L-OAM) and transverse OAM (T-OAM). Light carrying time-varying L-OAM, known as self-torqued light, was recently discovered during harmonic generation and has been extensively developed within the context of optical frequency combs (OFCs). Meanwhile, ultrafast bursts of optical pulses, analogous to OFCs, are sought for various light-matter interaction, spectroscopic and nonlinear applications[1-6]. However, achieving transiently switchable T-OAM of light on request, namely spatiotemporal vortex pulse bursts, with independently controlled spatiotemporal profile of each comb tooth, remain unrealized thus far. In this work, the experimental generation of spatiotemporal vortex bursts featured with controllable time-dependent characteristics is reported. The resultant bursts comprised of spatiotemporal optical vortex comb teeth have picosecond timescale switchable T-OAMs with defined arrangement, manifesting as spatiotemporal torquing of light. We also show ultrafast control of T-OAM chirality, yielding pulse bursts with staggered azimuthal local momentum density, resembling Kármán vortex streets. This approach enables the tailoring of more intricate spatiotemporal wavepacket bursts, such as high-purity modes variation in both radial and azimuthal quantum numbers of spatiotemporal Laguerre-Gaussian wavepackets over time, which may facilitate a host of novel applications in ultrafast light-mater interactions, high-dimensional quantum entanglements, space-time photonic topologies as well as spatiotemporal metrology and photography.**

## Introduction

Finite trains of ultrashort pulses, commonly referred to as pulse bursts and similar to optical frequency combs consisting of a series of separated "comb teeth" of frequencies, have already been utilized in various applications, including laser materials processing[1,2], nonlinear science[3,4], ultrafast photography[5,6], among many others. In recent years, new research avenues have been opened up by exploring two-dimensional spatiotemporally structured light with desired attributes. Spatiotemporally coupled wavepackets with diverse geometric and topological textures exhibit unique physical properties during propagation, focusing, and interaction with matter[7]. However, versatile pulse bursts comprising spatiotemporal wavepackets with tailored photon degrees of freedom (DoFs) have not been reported[8,9]. Orbital angular momentum (OAM) as a promising DoF of light is associated with its twisted waveform, resulting in phase singularities surrounded by a symmetric doughnut intensity profile[10]. The photonic intrinsic (coordinates-independent) OAM can either be longitudinal (L-OAM) if the spiral phase in the spatial domain or transverse (T-OAM) in the space-time domain[11-13]. The L-OAM was early discovered in spatial optical vortices by Allen et al. in 1992[14], which has been extensively studied and has spurred numerous significant applications[15,16]. The T-OAM is a newly discovered property of photons in spatiotemporal optical vortices (STOVs) in recent experimental demonstrations[17,18]. This revolutionary progress propels research towards nonlinear optics[19-21], novel photonics topology[22,23], spatiotemporal wavepackets engineering[24-26] and other matter waves[27-29], etc. Spatiotemporal structure wavepackets featuring controllable multiple photon DoFs have also been experimentally constructed[30]. While a conventional spatial/spatiotemporal vortex beam generally carries a temporally static OAM. In this context, self-torqued beam with time-varying OAM was firstly created in high harmonic generation driven by two ultrafast pulses with different OAM and time delayed to each other[31]. Since then, spatiotemporal wavepackets carrying time-dependent OAM have been sequentially synthesized, remarkably in ultrafast light coils through correlating spatial modes with time frequencies[32-35], tailored spatiotemporal dynamic wavepackets by means of a multi-plane light conversion[36,37], integrated vortex emitter by imprinting OAM modes into frequency microcomb[38,39]. However, to date, the majority of self-torqued light are primarily restricted to helical wavepackets, that is time-varying L-OAMs along light pulse, as shown in Fig. 1a.

Very recently, the concept of spatiotemporal self-torqued light holds time-varying T-OAM gradually emerged. Particularly by imparting a transient phase perturbation into an STOV pulse, subtle changes in the quantity of T-OAM over time have been successfully observed in the experiments[40]. Through intentionally offsetting spiral phases along the frequency, the resulting "STOV strings" exhibit time separated singularities[41,42]. Despite these remarkable developments, they are not suitable for practical scenarios since the distorted STOV's profile not only seriously impairs the purity of T-OAM but also raises severely coupled crosstalks between singularities after transmission. Thus generation of a set of isolated spatiotemporal wavepackets with time-varying T-OAM remains to be demonstrated. In this work, we achieve an ultrafast pulse train of independent spatiotemporal Laguerre-Gaussian (STLG) basis with variable quantum numbers, constructing "spatiotemporal vortex bursts" with tailored time-varying DoFs (Fig. 1b), that resembles OFCs. Instead of conventional spatial modes[36,37], each comb tooth within the burst lies on the space-time plane in a spatiotemporally coupled manner. The magnitude and arrangement of radial and azimuthal quantum numbers printed on each comb tooth can be self-defined on demand. The resulted spatiotemporal vortex bursts appear to resemble flying donuts with ultrafast switchable quantum numbers, reaching terahertz intraburst repetition rate conversion between high-purity mode indices.

**Principle and Experiment**

The conceptual scheme for the synthesis of spatiotemporal vortex bursts is illustrated in Fig. 1c,d. Optical pulse bursts and frequency combs can be viewed as inverse processes in the time-frequency domain. By correlating spatial vortices with finite comb tooth within an OFC, self-torqued helical pulse with time-varying L-OAMs can be generated in the time domain[38,39]. Conversely, an optical pulse burst exhibits a pulse comb in the time domain resulting in a complex spectrum like the temporal representation of such OFC (Methods), as shown in Fig. 1c. Therefore, to synthesize a pulse burst comprising a chain of regular spatiotemporal vortex pulses, we assign specific spatiotemporal coupling mode into each comb teeth in the time domain, forming a complex 2D spatial-spectral representation (Fig. 1d). Such spatial-spectral representation for spatiotemporal tailored wavepackets can be realized by the integration of a pulse shaper and a complex modulation holography[30] (which is termed spatiotemporal multiplexing hologram here, Methods). In the experiment, the hologram mask displayed on a reflective phase-only spatial light modulator is inserted in the intermediate confocal plane of a 4f ultrafast pulse shaper. With the help of a grating and a cylindrical lens, the spatial-spectral distribution of an incident pulse can be readily tailored by this mask. Since all of comb teeth within a burst are spatiotemporally coherent, we can use a single Fourier transform-limited pulsed reference beam to interrogate their 3D spatiotemporal morphology (see Supplementary Note 1 for the detailed experiment principle). The intraburst repetition rate and order inside the synthesized spatiotemporal vortex burst are determined by the intraburst phase slips (Methods). In addition to imparting the T-OAM content, a spectral group delay dispersion phase is added to chirp the pulse for balancing spatial diffraction and temporal dispersion, which is essential to produce a high mode purity spatiotemporal vortex burst (Supplementary Note 2).

**Results**

We firstly generate a spatiotemporal vortex burst with time-varying azimuthal quantum numbers from $\ell$ = -3 to +3 in Fig. 2. Such spatiotemporal vortex burst is designed as consisting of 7 LG bases of radial index $p$ = 0 with a temporal pitch of 2.8ps. Fig. 2a presents the 3D iso-intensity profile of the synthesized burst in the space-time domain, reconstructed from 600 temporal slices from 0ps to ~20ps in the experiment (Supplementary Note 1 and Supplementary Movie 1). The measured iso-surface showcases a pulse comb consisting of a series of STOV with variable sizes, each separated from the adjacent ones by 2.8ps. Each comb tooth exhibits a central dark hole induced by their spatiotemporal phase singularities, as shown in Fig. 2b. The phase distribution also confirms the time-varying azimuthal quantum numbers from $\ell$ = -3 to +3. The T-OAM spectrum power distributions (Supplementary Note 3) of the synthesized spatiotemporal vortex burst are shown in Fig. 2c. The measured results present a temporal evolution of T-OAM and clearly confirm the spatiotemporal torquing of light (appears an angular acceleration/deceleration on the space-time plane, as plotted in Fig. 2d), which implies a terahertz variation $\partial l(\theta_{x-t})/\partial t \sim 0.36$ THz of T-OAMs.

The proposed scheme enables easy customization of spatiotemporal vortex bursts with self-defined arrangement of T-OAMs. To illustrate this capacity, we generate a spatiotemporal vortex burst involving a series of spatiotemporal vortices with alternating azimuthal quantum numbers between $\ell = +1$ and $\ell = -1$. Fig. 3a shows the measured 3D iso-intensity (at 11% of the peak intensity) profile of this spatiotemporal vortex burst, which includes 5 STOVs with 2.8ps temporal pitch within a total temporal span of 12ps. Each comb tooth features an independent dark hole, resulting in a flying doughnut-shaped appearance. The spatiotemporal evolution of intensity and phase of the bursts can be easily demonstrated by their time-dependent interfered fringes (Supplementary Note 4). Fig. 3b presents the phase distribution of the generated burst, showing the temporally alternating variation of $\ell$ between +1 and -1 within this burst. To further elucidate the local spatiotemporal distributions of optical energy inside the burst, we plot the complex field and the canonical momentum density fluxes (Supplementary Note 5) in Fig. 3c. The top inserts display the corresponding calculated T-OAM per photon carried by each comb teeth. All results indicate the staggered reversal of energy flux chirality over time, forming an optical analogy to the Kármán vortex street (KVS). This phenomenon, characterized by the alternating chirality of swirling vortices, is caused by the nonlinear process of vortex shedding in fluid dynamics and has also been predicted in the linear realm[43] (see Supplementary Note 6). The instantaneous temporal variation of the T-OAM in a pulse burst is expected to play a role in the light-matter interactions, unveiling their energy transfer[44,45]. We can actively adjust the temporal separation and order of azimuthal quantum numbers within a burst, displaying self-defined time-varying T-OAMs (Supplementary Note 7).

Beyond the time-dependent azimuthal quantum numbers shown, our scheme can be readily reconfigured on demand to create more elaborate spatiotemporal dynamics. To further illustrate this versatility, we generate a spatiotemporal wavepacket burst involving three STLG wavepackets, a set of complete and orthonormal basis, each assigned to a distinct azimuthal ($\ell$) and radial ($p$) quantum number. Fig. 4a shows the measured 3D iso-intensity profile of the synthesized STLG wavepacket burst, which involves 3 STLG wavepackets ($p = 1 \& l = -1$, $p = 2 \& l = +1$ and $p = 1 \& l = +2$) with 4ps temporal pitch within a temporal span of 12ps. The corresponding spatiotemporal phase distribution is plotted in Fig. 4b. This burst shows dynamic spatiotemporal phase singularities and edge dislocations along time, resulting in time-varying coupled multi-ring topologies. Fig. 4c shows the modal analysis for each comb tooth of this burst (Supplementary Note 8). The power distribution located at the desired radial and azimuthal index is examined to be 82.5%, 52.34 and 84.4% respectively at the time $t_0 \approx 2$ps, 6ps and 10ps, clearly confirming the generation of high-purity time-varying STLG wavepackets. This involves carving dual time-varying DoFs of light field in a pulse burst. Additionally, we can achieve the spatiotemporal collision of dual STLG wavepackets inside a burst by controlling the relative delay between comb teeth, leading to more complex time-varying spatiotemporal properties (Supplementary Note 9).

## Discussions

In summary, we have demonstrated a new type of spatiotemporal vortex bursts featured by tailored spatiotemporal coupling properties at variable temporal locations. The generated spatiotemporal vortex bursts consist of tailored spatiotemporal vortex wavepackets, exhibiting temporally varied T-OAMs at picosecond timescale, manifesting as self-torque of spatiotemporal light. Our scheme achieves temporal variation with self-defined arrangements on both the radial and azimuthal quantum numbers of STLG wavepackets. Our results enhance the ability to precisely control ultrashort pulses in both time and space simultaneously. The methodology demonstrated here could be further developed to tailor more sophisticated spatiotemporal structure wavepackets with tunable dynamic properties[36,46] (Supplementary Note 11). The preparation of spatiotemporal vortex bursts is of great importance for revealing T-OAM-matter interactions in physicochemical processes, space-time photonic topologies, multiplexed optical communications and ultrafast events probing, etc. Although the bulky elements with a limited resolution in our demonstrations hampers the number of pulses per burst (Supplementary Note 12), employing advanced metasurface devices could further significantly enhance modulation efficiency, enable ultra-high resolution and ultra-high bandwidth for spatiotemporal light control[47,48].

## Methods

**Optical pulse burst and its spectral characteristics.** Let's start with a conventional 1D pulse burst characterized by invariant temporal separation, consistent across all pulses[8,9]. Neglecting spatial effects and using the slowly varying envelope approximation, an optical burst comprising $N$ pulses with electric fields can be described as follows:

$$E_n(t) = c_n(t)\exp[i\phi_n(t)], \tag{1}$$

where $c_n(t)$ and $\phi_n(t)$ is the amplitude and phase of the $n$th envelop. The pulse burst is defined in the time domain as

$$E(t) = \sum_{n=1}^{N} E_n(t - n\Delta\tau), \tag{2}$$

where $\Delta\tau$ being the intraburst temporal pitch. In our analysis, assuming all sub pulses are identical in amplitude, i.e. $E_n = E_p$ for $n = 1,2,\ldots,N$. The spectral representation of Eq. (2) is expressed in the frequency domain as

$$\tilde{E}(\omega) = \sum_{n=1}^{N} \tilde{E}_n(\omega) \exp(in\Delta\tau\omega) \tag{3}$$

$$= \tilde{E}_p(\omega) \cdot \frac{1+e^{-iN\Delta\tau\omega}}{1+e^{-i\Delta\tau\omega}}. \tag{4}$$

Eq. (3) and Eq. (4) give the spectral representations of conventional pulse bursts. The last term in Eq. (4) shows the spectral interference factor as displayed in Fig. 1c, for $N$=1, Eq. (4) reduces to a single pulse spectrum. We plot the formation of spectral modes of single pulse, $N$=4 Gaussian pulse bursts with/without chirps in the Supplementary Note 10.

**Spatiotemporal wavepackets bursts and spatiotemporal multiplexing holography.** To assign specific spatiotemporal coupling wavepackets to the comb teeth of a pulse burst, from Eq. (3), the equivalent spatial-spectral representation of a spatiotemporal vortex burst can be expressed as

$$E(\xi,\Omega) = \sum_{p,\ell} c_{p,\ell} LG_p^\ell(\xi,\Omega)\exp(i\,\Omega/f_{p,\ell}), \tag{5}$$

where $\Omega = \omega - \omega_0$ is the reduced angular frequency and $LG_p^\ell(\xi,\Omega)$ denotes Laguerre-Gaussian modes with the dual DoFs of the radial index $p$ and azimuthal index $\ell$[30]. $c_{p,\ell}$ can be identical or different and determines the weights of each comb teeth. The last term $\exp(i\,\Omega/f_{p,\ell})$ denotes the intraburst phase slip, which determinates the intraburst repetition rates. Note that such 2D spatial-spectral representation contains amplitude and phase information. As such, neglecting either the amplitude or the phase information results in strong crosstalks between each mode. Specifically, we adopt the following phase-only hologram to implement a complex-amplitude modulation[49]:

$$\psi_{\text{SLM}}(\xi,\Omega) = \text{mod}\{\psi_E + \pi\,\text{sinc}^{-1}|E| + g\xi\,\text{sinc}^{-1}[1-|E|] + \beta\Omega^2, 2\pi\}, \tag{6}$$

where $\psi_E$ stands for the phase of $E(\xi,\Omega)$ and $g$ denotes the frequency of a linear phase ramp, the depth of which is contingent upon the modulus of $E(\xi,\Omega)$. The coherent superposition of LG modes of Eq. (5) generates a spatiotemporal multiplexing hologram [Eq. (6)], each mode carrying its own spatiotemporal features is, in principle, distinguishable from the others in the time domain. As such, after a space-time Fourier transformation, the resultant field demultiplexed along time dimension forming a pulse burst of spatiotemporal vortex pulses (see Supplementary Note 1):

$$E(r_{x-\tau},\theta_{x-\tau}) = \sum_{p,\ell} c_{p,\ell} LG_p^\ell(r_{p,\ell},\theta_{p,\ell}), \tag{7}$$

where $r_{p,\ell} = \sqrt{(\tau - t_{p,\ell})^2 + \beta^2 x^2}$, $\theta_{p,\ell} = \tan^{-1}[\beta x/(\tau - t_{p,\ell})]$ and $t_{p,\ell} = 1/f_{p,\ell}$. A spatiotemporal vortex burst consisting of isolated LG orthogonal basis is a solution to the scalar wave equation, which performs self-similar evolutions and remains pure vortex modes under dispersion matching conidiation to avoid intermodal crosstalks[30].

### Acknowledgments

We acknowledge financial support from the National Key Research and Development Program of China (2022YFA1404800 [Y.C.], 2019YFA0705000 [Y.C.]), National Natural Science Foundation of China (92050202 [Q.Z.], 12192254 [Y.C.], 92250304 [Y.C.], 12374311 [C.L.]). C.L. also acknowledges support by the Taishan Scholars Program of Shandong Province (Grant No. tsqn202312163) and the Natural Science Foundation of Shandong Province (Grant No. ZR2023YQ006). Q.C. also acknowledges support by the Shanghai Sailing Program (Grant No. 21YF1431500). Q.Z. also acknowledges support by the Key Project of Westlake Institute for Optoelectronics (Grant No. 2023GD007).

### Author contributions

X.L. Y.C. and Q.Z. conceived the idea and initiated this project. X.L. designed and performed all the experiments and calculations. X.L., C.L. and Q.C. both analyzed the data. X.L. and Q.Z. prepared the manuscript. Y.C. and Q.Z. supervised the project. All authors contributed to the discussion and writing of the manuscript.

## Data availability

All other data are available in the article and its supplementary files or from the corresponding authors upon request.

## Conflict of interest

The authors declare no competing interest.

## Supplementary information

See Supplementary information file for supporting content.

## References

1. Kerse, C., Kalaycıoğlu, H., Elahi, P., Çetin, B., Kesim, D. K., Akçaalan, Ö., Yavaş, S., Aşık, M. D. Öktem, B., Hoogland, H. Holzwarth, R. & Ilday, F. Ö. Ablation-cooled material removal with ultrafast bursts of pulses. *Nature* **537**, 84-88 (2016).
2. Park, M., Gu, Y., Mao, X., Grigoropoulos, C. P., & Zorba, V. Mechanisms of ultrafast GHz burst fs laser ablation. *Sci. Adv.* **9**, eadf6397 (2023).
3. Hu, H., Flöry, T., Stummer, V., Pugzlys, A., Zeiler, M., Xie, X., Zheltikov, A. & Baltuška, A. Hyper spectral resolution stimulated Raman spectroscopy with amplified fs pulse bursts. *Light-Sci. Appl.* **13**, 61 (2024).
4. Umstadter, D., Esarey, E., & Kim, J. Nonlinear plasma waves resonantly driven by optimized laser pulse trains. *Phys. Rev. Lett.* **72**, 1224 (1994).
5. Nakagawa, K., Iwasaki, A., Oishi, Y., Horisaki, R., Tsukamoto, A., Nakamura, A., Hirosawa, K., Liao, H., Ushida, T., Goda, K., Kannari, F. & Sakuma, I. Sequentially timed all-optical mapping photography (STAMP). *Nat. Photonics* **8**, 695-700 (2014).
6. Liu, J., Marquez, M., Lai, Y., Ibrahim, H., Légaré, K., Lassonde, P., Liu, X., Hehn, M., Mangin, S., Malinowski, G., Li, Z., Légaré, F. & Liang, J. Swept coded aperture real-time femtophotography. *Nat. Commun.* **15**, 1589 (2024).
7. Zhan, Q. Spatiotemporal sculpturing of light: a tutorial. *Adv. Opt. Photon.* **16**, 163-228 (2024).
8. Cundiff, S. T. & Weiner, A. M. Optical arbitrary waveform generation. *Nat. Photonics* **4**, 760-766 (2010).
9. Stummer, V., Flöry, T., Krizsán, G., Polónyi, G., Kaksis, E., Pugžlys, A., Hebling, J., Fülöp, J. A. & Baltuška, A. Programmable generation of terahertz bursts in chirped-pulse laser amplification. *Optica* **7**, 1758-1763 (2020).
10. Padgett, M. & Bowman, R. Tweezers with a twist. *Nat. Photonics* **5**, 343–348 (2011).
11. Bliokh, K. Y. & Nori, F. Transverse and longitudinal angular momenta of light. *Phys. Rep.* **592**, 1–38 (2015).
12. Bliokh, K. Y. & Nori, F. Spatiotemporal vortex beams and angular momentum. *Phys. Rev. A* **86**, 033824 (2012).
13. Porras, M. A. Transverse orbital angular momentum of spatiotemporal optical vortices. *Prog. Electromagn. Res.* **177**, 95 (2023).

14. Allen, L., Beijersbergen, M. W., Spreeuw, R. J. C. & Woerdman, J. P. Orbital angular momentum of light and the transformation of Laguerre-Gaussian laser modes. *Phys. Rev. A* **45**, 8185-8189 (1992).
15. Yao, A. M. & Padgett, M.J. Orbital angular momentum: origins, behavior and applications. *Adv. Opt. Photon.* 3, 161-204 (2011).
16. Shen, Y., Wang, X., Xie, Z., Min, C., Liu, Q., Gong, M. & Yuan, X. Optical vortices 30 years on: OAM manipulation from topological charge to multiple singularities. *Light Sci. Appl.* **8**, 90 (2019).
17. Hancock, S. W., Zahedpour, S., Goffin, A. & Milchberg, H. M. Free-space propagation of spatiotemporal optical vortices. *Optica* **6**, 1547-1553 (2019).
18. Chong, A., Wan, C., Chen, J. & Zhan, Q. Generation of spatiotemporal optical vortices with controllable transverse orbital angular momentum. *Nat. Photonics* **14**, 350-354 (2020).
19. Gui, G., Brooks, N. J., Kapteyn, H. C., Murnane, M. M. & Liao, C. T. Second-harmonic generation and the conservation of spatiotemporal orbital angular momentum of light. *Nat. Photonics* **15**, 608-613 (2021).
20. Hancock, S. W., Zahedpour, S. & Milchberg, H. M. Second-harmonic generation of spatiotemporal optical vortices and conservation of orbital angular momentum. *Optica* **8**, 594-597 (2021).
21. Fang, Y., Lu, S. & Liu, Y. Controlling photon transverse orbital angular momentum in high harmonic generation. *Phys. Rev. Lett.* **127**, 273901 (2021).
22. Wan, C., Shen, Y., Chong, A. & Zhan, Q. Scalar optical hopfions. *eLight* **2**, 1-7 (2022).
23. Shen, Y., Zhang, Q., Shi, P., Du, L., Yuan, X. & Zayats, A.V. Optical skyrmions and other topological quasiparticles of light. *Nat. Photonics* **18**, 15–25 (2024).
24. Wan, C., Cao, Q., Chen, J., Chong, A. & Zhan, Q. Toroidal vortices of light. *Nat. Photonics* **16**, 519-522 (2022).
25. Cao, Q., Chen, J., Lu, K., Wan, C., Chong, A. & Zhan, Q. Non-spreading Bessel spatiotemporal optical vortices. *Sci. Bull.* **67**, 133-140 (2022).
26. Chen, W., Zhang, W., Liu, Y., Meng, F. C., Dudley, J. M. & Lu, Y. Q. Time diffraction-free transverse orbital angular momentum beams. *Nat. Commun.* **13**, 4021 (2022).
27. Ge, H., Liu, S., Xu, X., Long, Z., Tian, Y., Liu, X., Lu, M. & Chen, Y. Spatiotemporal Acoustic Vortex Beams with Transverse Orbital Angular Momentum. *Phys. Rev. Lett.* **131**, 014001 (2023).
28. Zhang, H., Sun, Y., Huang, J., Wu, B., Yang, Z., Bliokh, K. Y. & Ruan, Z. Topologically crafted spatiotemporal vortices in acoustics. *Nat. Commun.* **14**, 6238 (2023).
29. Bliokh, K. Y. Orbital angular momentum of optical, acoustic, and quantum-mechanical spatiotemporal vortex pulses. *Phys. Rev. A* **107**, L031501 (2023).
30. Liu, X., Cao, Q., Zhang, N., Chong, A., Cai, Y. & Zhan, Q. Spatiotemporal optical vortices with controllable radial and azimuthal quantum numbers. *Nat. Commun.* **15**, 5435 (2024).
31. Rego, L., Dorney, K. M., Brooks, N. J., Nguyen, Q. L., Liao, C. T., San Román, J., Couch, D. E., Liu, A., Pisanty, E., Lewenstein, M. & Plaja, L. Generation of extreme-ultraviolet beams with time-varying orbital angular momentum. *Science* **364**, eaaw9486 (2019).

32. Zhao, Z., Song, H., Zhang, R., Pang, K., Liu, C., Song, H., Almaiman, A., Manukyan, K., Zhou, H., Lynn, B., Boyd, R.W., Tur, M. & Willner, A.E. Dynamic spatiotemporal beams that combine two independent and controllable orbital-angular-momenta using multiple optical-frequency-comb lines. *Nat. Commun.* **11**, 4099 (2020).
33. Chen, L., Zhu, W., Huo, P., Song, J., Lezec, H. J., Xu, T. & Agrawal, A. Synthesizing ultrafast optical pulses with arbitrary spatiotemporal control. *Sci. Adv.* **8**, eabq8314 (2022).
34. Piccardo, M., Oliveira, M., Policht, V. R., Russo, M., Ardini, B., Corti, M., Valentini, G., Vieira, J., Manzoni, C., Cerullo, G. & Ambrosio, A. Broadband control of topological–spectral correlations in space–time beams. *Nat. Photonics* **17**, 822-828 (2023).
35. Oliveira, M. & Ambrosio, A. Sub-cycle modulation of light's Orbital Angular Momentum. arXiv:2405.13723 (2024).
36. Cruz-Delgado, D., Yerolatsitis, S., Fontaine, N. K., Christodoulides, D. N., Amezcua-Correa, R. & Bandres, M. A. Synthesis of ultrafast wavepackets with tailored spatiotemporal properties. *Nat. Photonics* **16**, 686-691 (2022).
37. Cruz-Delgado, D., Antonio-Lopez, J.E., Perez-Leija, A., Fontaine, N.K., Eikenberry, S.S., Christodoulides, D.N., Bandres, M.A. & Amezcua-Correa, R. Spatiotemporal Control of Ultrafast Pulses in Multimode Optical Fibers. arXiv:2402.11783 (2024).
38. Chen, B., Zhou, Y., Liu, Y., Ye, C., Cao, Q., Huang, P., Kim, C., Zheng, Y., Oxenløwe, L., Yvind, K., Li, J., Li, J., Zhang, Y., Dong, C., Fu, S., Zhan, Q., Wang, X., Pu, M., & Liu, J. Integrated optical vortex microcomb. *Nat. Photonics* (2024).
39. Liu, Y., Lao, C., Wang, M., Cheng, Y., Wang, Y., Fu, S., Gao, C., Wang, J., Li, B. B., Gong, Q., Xiao, Y., Liu, W. & Yang, Q. F. Integrated vortex soliton microcombs. *Nat. Photonics* (2024).
40. Hancock, S. W., Zahedpour, S., Goffin, A. & Milchberg, H. M. Spatiotemporal Torquing of Light. *Phys. Rev. X* **14**, 011031 (2024).
41. Wan, C., Chen, J., Chong, A. & Zhan, Q. Generation of ultrafast spatiotemporal wave packet embedded with time varying orbital angular momentum. *Sci. Bull.* **65**, 1334–1336 (2020).
42. Huang, S., Li, Z., Li, J., Zhang, N., Lu, X., Dorfman, K., Liu, J. & Yao, J. Spatiotemporal vortex strings. *Sci. Adv.* **10**, eadn6206 (2024).
43. Shen, Y., Papasimakis, N. & Zheludev, N.I. Nondiffracting supertoroidal pulses and optical "Kármán vortex streets". *Nat. Commun.* **15**, 4863 (2024).
44. Hu, Y., Kingsley-Smith, J.J., Nikkhou, M., Sabin, J.A., Rodríguez-Fortuño, F.J., Xu, X. & Millen, J. Structured transverse orbital angular momentum probed by a levitated optomechanical sensor. *Nat. Commun.* **14**, 2638 (2023).
45. Ouyang, X., Xu, Y., Xian, M., Feng, Z., Zhu, L., Cao, Y., Lan, S., Guan, B.O., Qiu, C.W., Gu, M. & Li, X. Synthetic helical dichroism for six-dimensional optical orbital angular momentum multiplexing. *Nat. Photonics* **15**, 901–907 (2021).
46. Pierce, J.R., Palastro, J.P., Li, F., Malaca, B., Ramsey, D., Vieira, J., Weichman, K. & Mori, W.B. Arbitrarily structured laser pulses. *Phys. Rev. Research* **5**, 013085 (2023).

47. Shaltout, A. M., Lagoudakis, K. G., van de Groep, J., Kim, S. J., Vučković, J., Shalaev, V. M. & Brongersma, M. L. Spatiotemporal light control with frequency-gradient metasurfaces. *Science*, **365**, 374-377 (2019).
48. Shaltout, A. M., Shalaev, V. M. & Brongersma, M. L. Spatiotemporal light control with active metasurfaces. *Science* **364**, eaat3100 (2019).
49. Cao, Q., Zhang, N., Chong, A. & Zhan, Q. Spatiotemporal Hologram. arXiv:2401.12642 (2024).

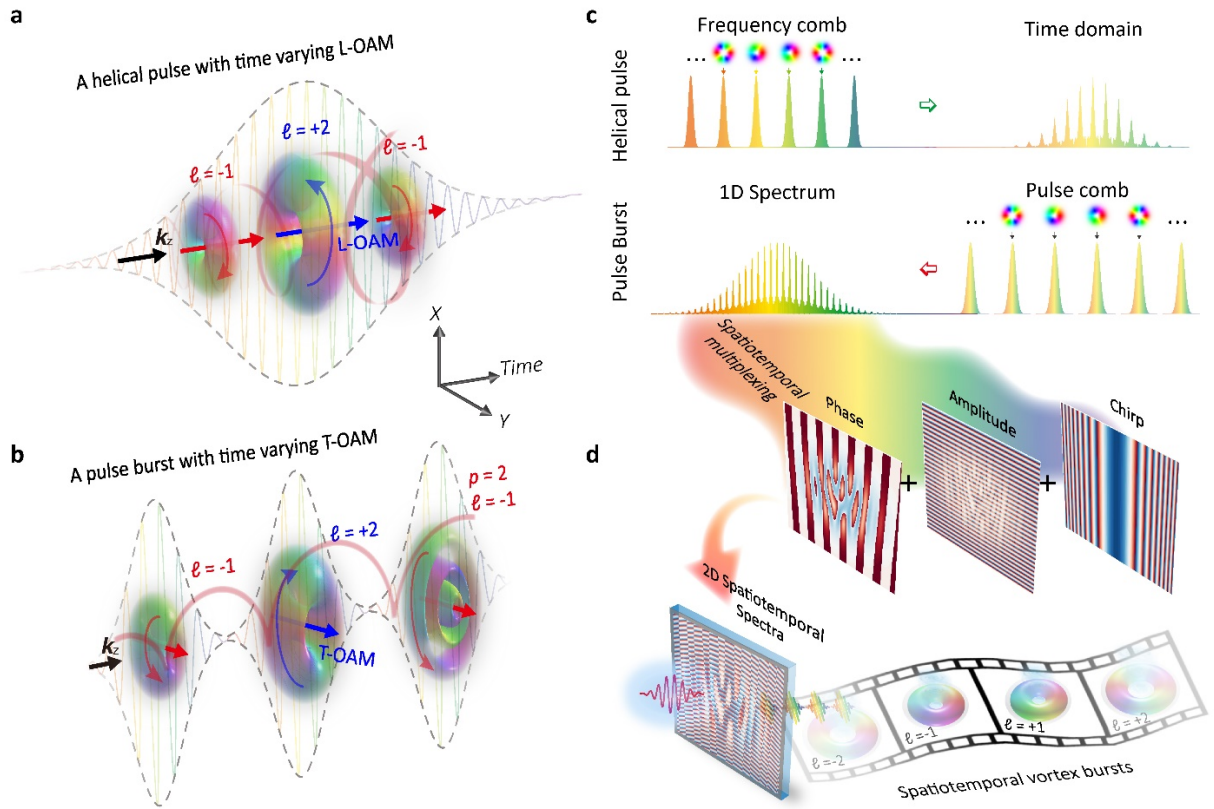

**Fig. 1 | Conceptual scheme of the synthesis of spatiotemporal vortex bursts with time-varying dynamic properties. a,** Time-varying L-OAMs of helical pulse. **b,** Time-varying T-OAMs of spatiotemporal vortex bursts. **c,** Analogy between optical frequency combs and pulse bursts. **d,** Principle for spatiotemporal vortex bursts generation by engineering the 2D spatiotemporal spectra (Methods).

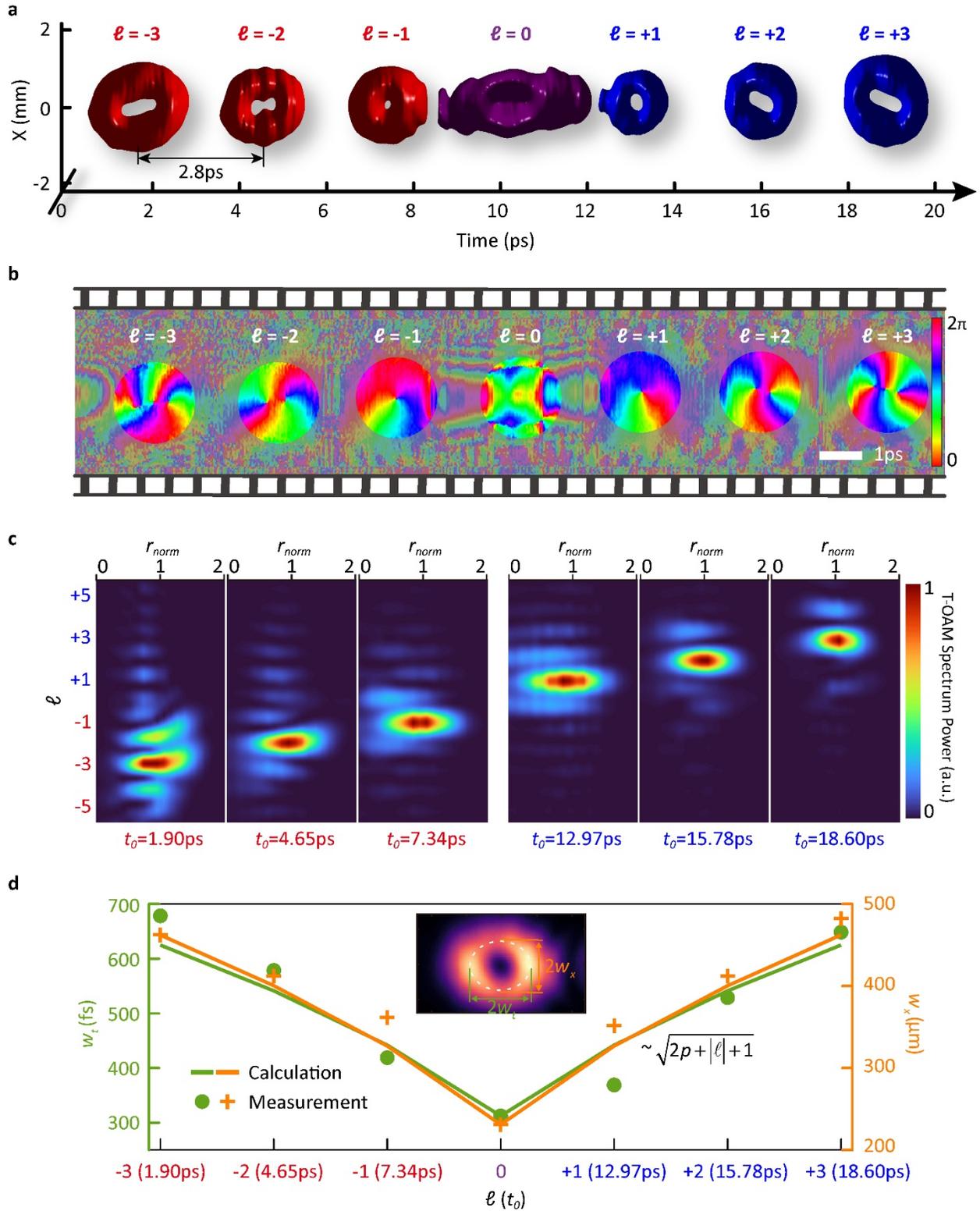

**Fig. 2 | Synthesized spatiotemporal vortex burst with time-varying T-OAM from $\ell$ = -3 to +3. a,** Measured iso-intensity profile of the spatiotemporal vortex burst in the space-time domain, which has a ~20ps temporal length and a 2.8ps temporal pitch. The isovalue is set to 8.5% of the

peak intensity. The red and blue color denotes the negative and positive $\ell$ respectively. **b,** Retrieved phase distribution in the meridional plane. **c,** T-OAM modal analysis of the synthesized spatiotemporal vortex bursts, exhibiting a spatiotemporal self-torque of light, where $r_{\text{norm}} = \sqrt{(T-t_0)^2/w_t^2 + (X-x_0)^2/w_x^2}$ is a normalized spatiotemporal polar radius (Supplementary Note 3). **d,** Measured temporal/spatial radius ($w_t$ and $w_x$) as a function of $\ell$ ($t_0$), showing a spatiotemporal angular deceleration and acceleration along time, indicating an angular deceleration/acceleration on the space-time plane.

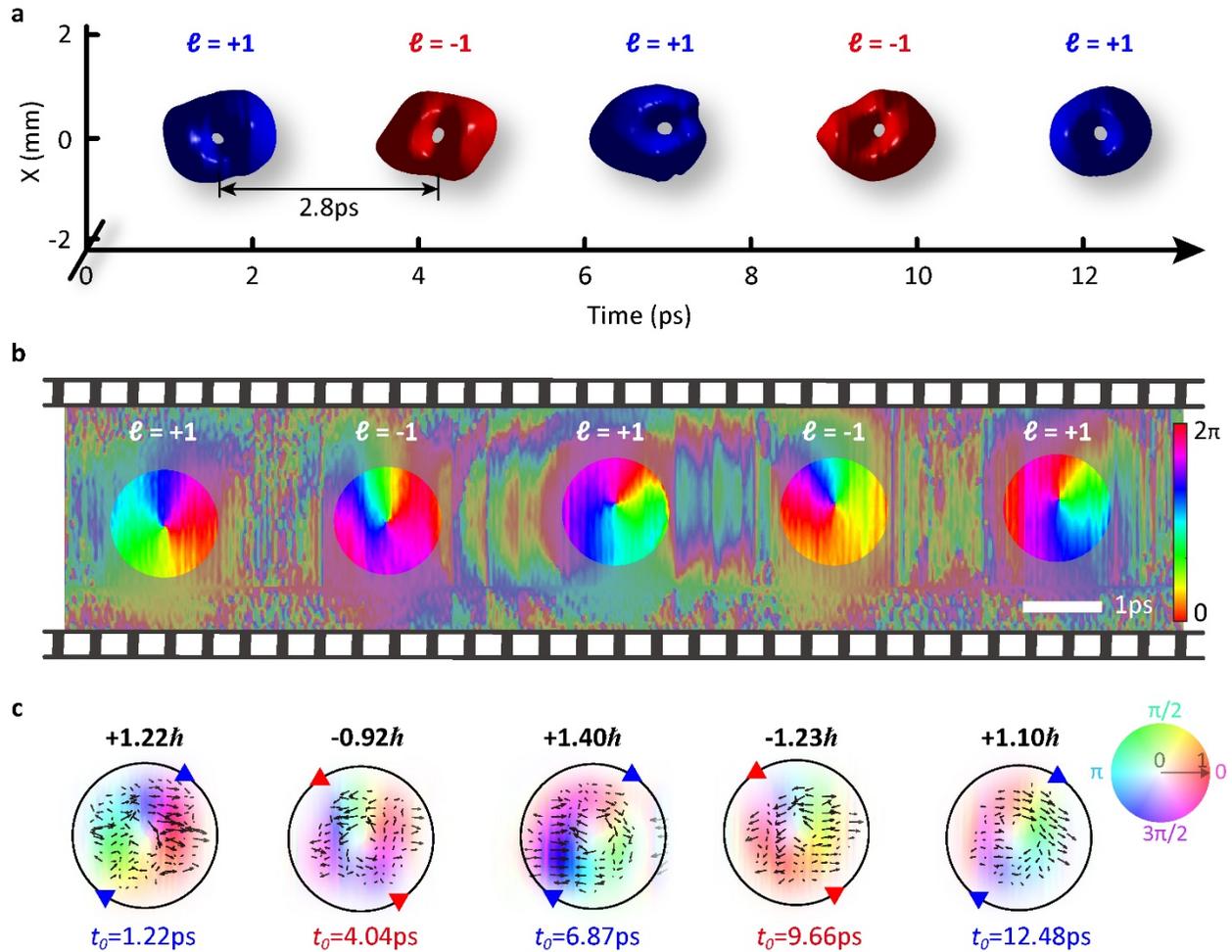

**Fig. 3 | Burst of spatiotemporal vortex pulses with time-varying chirality alternating between $\ell$ = +1 and $\ell$ = -1. a,** Measured iso-intensity profile. The isovalue is set to 11% of the peak intensity. **b,** Retrieved phase in the meridional plane. **c,** Extracted complex field and local momentum density flux distribution in the space-time domain, reminiscent of the KVS featured temporal variation of chirality of T-OAM. The saturation and hue denote amplitude and phase information respectively.

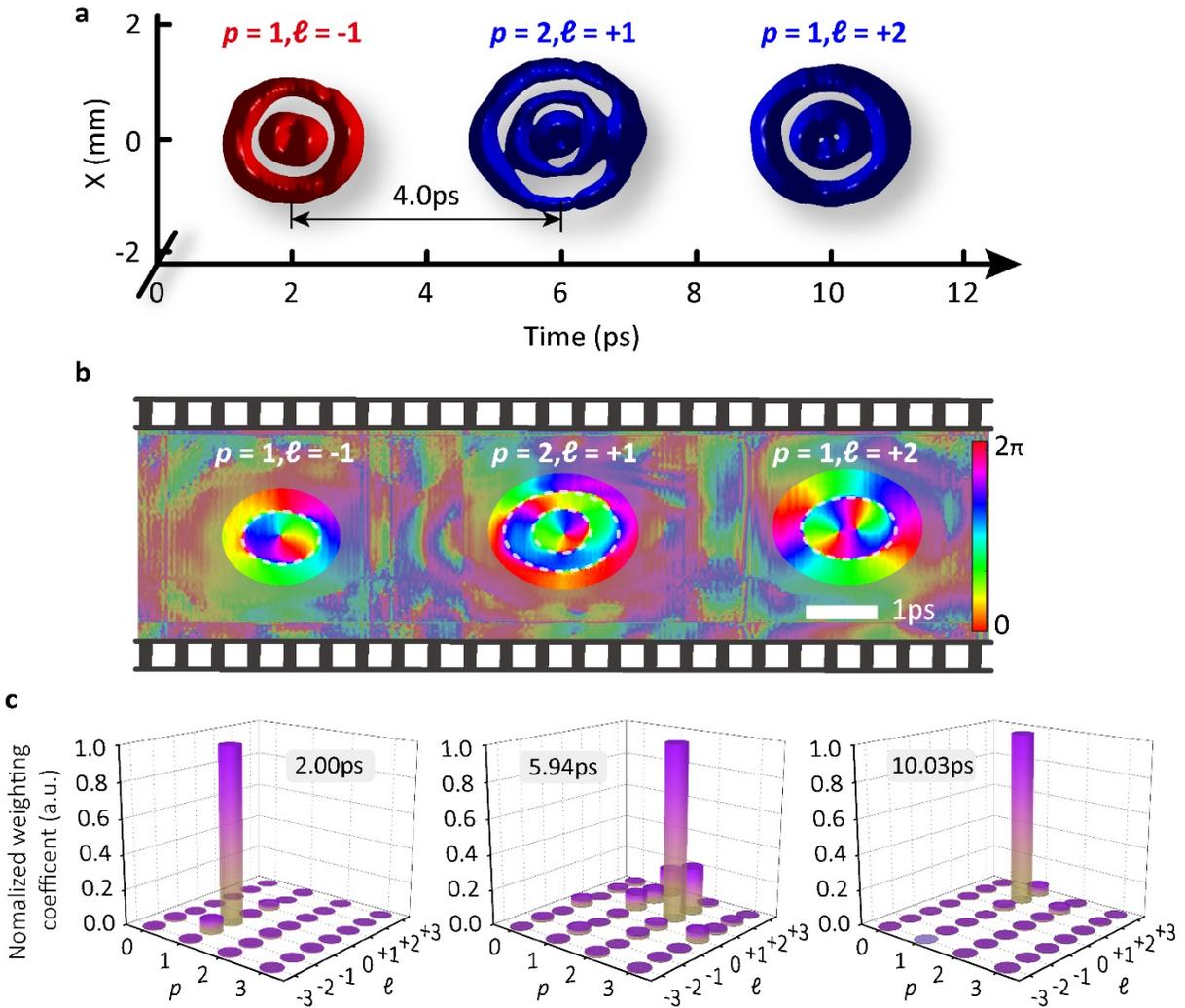

**Fig. 4 | Burst of spatiotemporal vortex pulses with defined time-varying quantum numbers on both radial and azimuthal indices. a,** Measured iso-intensity profile of this spatiotemporal vortex bursts, exhibiting a train of time-varying STLG wavepackets with variable $p$ and $\ell$. The temporal separation is 4ps. The isovalue is set to 5% of peak intensity. **b,** Retrieved phase distribution in the meridional plane, showing time-varying spatiotemporal phase singularities and edge dislocations. **c,** Modal weight analysis of each comb teeth at specific times within the burst, the power distributing at the desired radial ($p$) and azimuthal ($\ell$) index is 82.5%, 51.34% and 84.4%, respectively.